# Encapsulation, compensation, and substitution of catalyst particles during continuous growth of carbon nanotubes

Rong Xiang[§], Guohua Luo, Weizhong Qian[*]，Qiang Zhang, Yao Wang, Fei Wei[*]

*Beijing Key Laboratory of Green Chemical Reaction Engineering and Technology, Department of Chemical Engineering, Tsinghua University, Beijing 100084, P. R. China*

Qi Li

*Cnano Group Limited, 302 Hennessy Rd., Wanchai, Hong Kong, P. R. China*

Anyuan Cao

*Department of Mechanical Engineering, University of Hawaii at Manoa, Honolulu, HI 96822, USA*

[*] To whom correspondence may be addressed. Weizhong Qian, qwz@flotu.org, tel, +86-10-62788984, Weifei, weifei@flotu.org, tel, +86-10-62785464, Department of Chemical Engineering, Tsinghua University, Beijing 100084, China. Fax: +86-10-62772051

[§] Current address: Department of Mechanical Engineering, The University of Tokyo, 7-3-1 Hongo, Bunkyo-ku, Tokyo 113-8656, JAPAN.

A carbon nanotube (CNT) forms via incorporation of carbon into a metal particle (as catalyst) and subsequent precipitation into a tubular structure, in catalytic chemical vapor deposition (CVD).[1] The presence of this catalyst particle is essential for maintaining the precipitation process thus enabling continuous growth of the CNT. A unique property is that the CNTs are capable of drawing materials into their hollow cavities through open ends, resulting in the encapsulation of catalyst into CNTs.[2] Metal particles enclosed inside CNTs were frequently observed in samples made by CVD methods,[3,4] and nanowires with various compositions (Fe, Fe-Ni and Fe-Co alloy) have been made by choosing and combining catalyst sources, with hollow CNTs as templates.[5-7]

Continuous feeding of carbon source and catalyst precursor (e.g. ferrocene dissolved in xylene) is a widely used approach to produce long aligned CNTs with length up to millimeter scale.[4,8] Removal of ferrocene in the middle of CVD process resulted in substantial decrease of CNT growth rate or termination of growth in a short time,[9,10] whereas an extra addition of ferrocene during CVD was found to significantly boost the CNT length.[11] These results imply that catalyst particles may lose activity after a certain period, thus a continuous in situ refreshment of Fe catalyst is a crucial factor for sustaining CNT growth. Nonetheless, investigations on the mechanism have led to elusive results, with a number of models proposed in recent time.[4,12-14]

One of the key issues in exploring the growth mechanism is to understand the location of catalyst particles where incorporation of carbon occurs. Isotope labeling by $^{13}C$ in the carbon source has been used to detect the sequence of formation of different CNT sections.[15] Several recent reports have demonstrated "bottom-up growth" by lift-up stacking of multiple CNT layers. These results have proved a root growth mode for CNT formation in their CVD methods, in which catalyst particles stay at the bottom and CNTs extruding from the particles grow upward.[11,16,17]

As carbon nanotubes are capable of drawing materials inside,[3-7] we have explored this basic idea to track the distribution of catalyst particles by sequentially feeding two different catalyst sources (e.g. Ni and Fe). We found that catalyst materials at the ends of CNTs can be encapsulated into the CNT cavities constantly. However, such consumption of catalyst due to encapsulation could be balanced by continuous supply of additional catalyst source such that the CNTs maintain their growth process. The encapsulation of catalyst by CNTs occurs frequently (at an average period of several seconds), and was observed in virtually all the nanotubes which participated the growth. A CNT could uptake different materials when the catalyst precursors were switched between Ni and Fe during CVD. The self-renewal of catalyst due to constant consumption (encapsulation) and addition, as described here, is in contrast with previous understanding where catalyst particles were believed to stay permanently at the tip (or bottom) of nanotubes.

Figure 1a (top-left inset) shows the scanning electron microscopy (SEM) image of an as-grown CNT film produced by injection of nickelocene solution for 30 minutes followed by additional 30 minutes of ferrocene solution (see Experimental). Two separate reference experiments have shown that CNTs grew to 300 to 600 μm in length when only nickelocene was fed for half an hour, and 0.8 to 1 mm in length when ferrocene was fed alone. Here the total length of the film has reached 1.5 mm, showing that both nickelocene and ferrocene have effectively contributed to the growth of CNTs even they were introduced during different stages, and there was not much change of growth rate. Further, there is no obvious gap/interface present across the film section, indicating CNT growth was not interrupted to form two layers although catalyst precursors were switched during the CVD process.

A thin bundle of CNTs was peeled off from the film section to allow easy characterization by transmission electron microscopy (TEM) without disturbing their alignment, so that the top and bottom portions of film could be distinguished under TEM. Here we define the bottom of CNT films to be the portion close to the underlying substrate where CNTs grew. Following an individual CNT near the edge of the bundle, we observed large amounts of short catalyst rods (length of 50 to 100 nm) existing periodically along the CNT length at an average distance of 3 µm (Fig. 1). Scanning over other nanotubes and regions in the bundle showed encapsulated catalyst particles with comparable size and inter-particle distance. This shows that a substantial amount of catalyst materials supplied during CVD have been encapsulated into CNTs.

To track the distribution of Fe and Ni particles along the length of CNTs, we performed high resolution TEM and electron dispersive X-ray spectroscopy (EDS) on particles encapsulated at different positions of the CNT film (e.g. top, middle, and bottom) (Fig. 2a). Most of the particles near the top portion consist exclusively of Ni with a typical structure of Ni (FCC), and near the bottom part we found only γ-Fe (FCC) particles (Fig. 2b, 2c). The interface where the content of particles change from Ni to Fe was determined to be about 0.5 mm distance to the top surface, suggesting that nickelocene has catalyzed the growth of 1/3 of CNT length (~0.5 mm), while the remaining 2/3 of CNT length (~1 mm) was grown by ferrocene. This is reasonable considering the lower growth rate of CNTs by nickelocene in our CVD conditions (which is equal to about a half of the growth rate by ferrocene).

The distinct compositions of encapsulated particles in top or bottom parts of CNT films indicate two key features of the growth process. First, the bottom ends of CNTs are active growth sites where catalyst particles locate and sustain CNT growth by uptaking carbon source. In the first 30 minutes period, Ni particles (injected first) were encapsulated into CNT channels and lifted up with the growing CNTs, therefore finally they stay near the top portion of the film. In the second 30 minutes, Fe particles (injected afterwards) also entered CNTs, but their location is under earlier encapsulated Ni particles (Fig. 3a). Only when the encapsulation occurred at the bottom ends of CNTs, the resulted distribution is Ni on top of Fe. Filling from the top tips of CNTs will result in a reversed configuration, that is, Fe on top of Ni, as illustrated in Fig. 3b.

Second, the catalyst particles are not sitting at the ends of CNTs statically. Instead, they kept being drawn into CNT cavities, moving upward with growing CNTs. The consumption of catalyst pool for each CNT is very fast, although it is compensated by continuous supply of catalyst source. Based on an average particle separation of 3 μm as shown in Fig. 1b, and a measured growth rate of CNTs of 25 μm/min (= 1.5 mm/60 min), the encapsulation of the next particle occurs within 7 seconds (in which CNTs grow about 3 μm in length). If we assume a CNT diameter of 50 nm with encapsulated catalyst rods 10 nm in diameter and 50 nm in length, and also that the catalyst pool at the CNT ends is in spherical shape with similar size to CNT (50 nm), the volume of one encapsulated rod, which is equal to $\pi(10/2)^2 \times 50$ nm$^3$, is about 6 vol% of the catalyst pool ($4/3 \times \pi(50/2)^3$). This indicates that without additional supply, the catalyst pool will dry out in 2 minutes after about 17 encapsulation events. This is in agreement with previous observation that once ferrocene was removed from the injected solution, the growth of CNTs terminated in couple of minutes resulting in a thin layer rather than aligned film.[10] Thus compensation of catalyst (to prevent total consumption) is critical for sustaining continuous growth of CNTs.

In addition, the self-renewal of catalyst particles is not limited to the same material, considering of the existence of both Ni and Fe in the same CNT. Fe can be readily dissolved into Ni at temperature of about 500 ºC,[18] thus replacing Ni particles at the CNT ends. No trace of Ni was detected at the bottom ends of CNTs under TEM. The exclusive distribution of Fe particles near the bottom portion indicates that Fe has completely replaced Ni initially fed as the catalyst, and served for CNT growth thereafter. Characterization of encapsulated particles near the transition region where Ni changed to Fe by TEM and EDS has revealed both Ni and Fe content in the same particle (Fig. S1), indicating that CNT growth was not disturbed by switching catalyst materials.

Moreover, we have observed the presence of a structure consisting of curved graphene caps transversely located in the CNT channel (Fig. 4a, b). Such graphene caps were initially formed on the catalyst surface at the bottom of CNTs and moved upward during growth,[19] thus usually they are oriented toward the top of film (Fig. 4c). Many catalyst particles were enclosed in a chamber sealed by two graphene caps on both sides (Fig. 4a inset). As iron can diffuse through graphite layers only at temperatures of above 1600 ºC,[20] far higher than our CVD condition (800 ºC), the graphene caps connected to CNT walls have made an envelope and the only entrance for metal particle is the bottom ends, which further proved that the ends of CNTs are active growth sites where catalyst particles reside.

We have performed CVD by injecting ferrocene for relatively long time (e.g. 2 hours) to see whether the CNT length has effect on the self-renewal behavior of catalyst. With increasing CVD time, the growth of CNTs has gradually slowed down, especially after a period of 1 hour (Fig. 4d). The length of CNTs reached about 1.5 mm in 1 hour, but only increase to 2 mm after 2 hours growth even at a constant rate supply of catalyst precursor. One possible explanation is that the diffusion of catalyst (and carbon source) becomes more difficult through a very thick film to reach the active growth site (catalyst pool at the bottom surface). Such gas-diffusion-controlled mode in base-growth of CNTs has been observed previously.[17] Therefore once the encapsulation of catalyst is faster than feeding in at prolonged process, it may cause total trapping of catalyst and termination of growth. In addition, we measured the diameter of CNTs at different regions along the film section. It shows that the Ni catalyzed part of CNTs have an outer diameter of 50 ~60 nm, slightly larger than the Fe-catalyzed part (about 40 nm). The inner diameter for both regions ranges from 6 to 10 nm. The distance between adjacent Fe particles is slightly larger than that of Ni particles (~3 μm). The less frequent encapsulation of Fe particles might due to more difficult diffusion of Fe through thick CNT films.

In summary, we reported direct experimental evidence that the bottom ends of CNTs are active growth sites, and catalyst particles experience a constant encapsulation by CNTs, but can be compensated by additional source (even with different materials) to sustain a continuous growth. Our results should be referential for further understanding the CNT formation mechanism, and to control the growth process. CNTs encapsulated with different materials should enable the study on their electronic or magnetic properties, with potential applications as building blocks for nanoelectronics, and as fillers in composites for electromagnetic shielding.

*Experimental*

Aligned CNT arrays were synthesized by CVD using metalocene as catalyst precursor and cyclohexane as the solvent and carbon source, as described previously.[8] The difference here is that two types of precursors, nickelocene and ferrocene, were introduced sequentially into the CVD system, each for a period of 30 minutes. We used two separate injection tubes for switching between nickelocene and ferrocene dissolved in cyclohexane, to prevent contamination by residue from previous solution. The switching step was done in a few seconds and did not cause interruption on CNT growth. Both solutions had a metalocene concentration of 20 g/L, and the feeding rate was controlled by a motorized syringe pump at 5 mL/hour. Ar gas (containing 10 % hydrogen) was flowing through the CVD furnace at 600 ml/min. The reaction temperature was set to be 800 ℃ for both nickelocene and ferrocene periods. Although the highest growth rate for Ni was at 750 ℃, we used 800 ℃ for both Fe and Ni to avoid changing and fluctuation of temperature during CVD process. CNTs were collected from a quartz sheet placed in the middle of furnace for characterization.

**Acknowledgement.** The work was supported by China National '863' Program (No. 2003AA302630), China National Program(No. 2006CB0N0702), NSFC Key Program (No. 20236020), FANEDD (No. 200548), Key Project of Chinese Ministry of Education (No. 106011), THSJZ, and National center for nanosicence and technology of China (Nanoctr).

**Figure captions:**

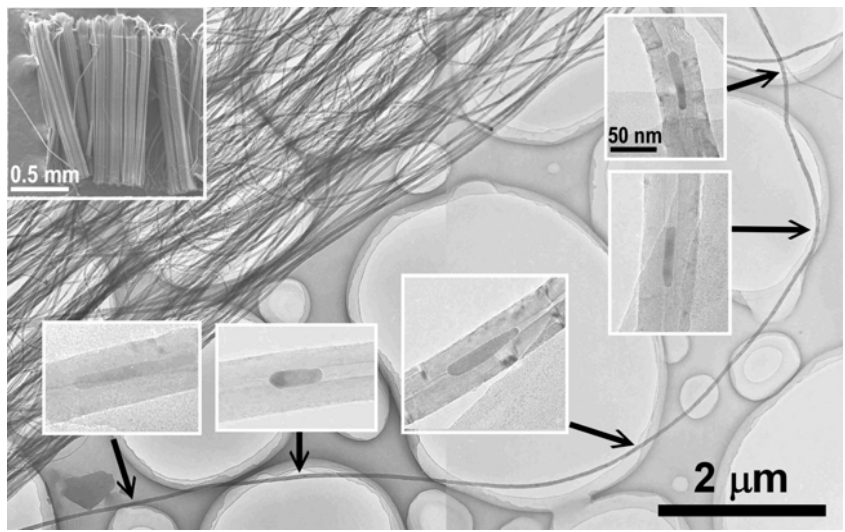

**Figure 1.** Characterization of CNTs and encapsulated catalyst particles. (top-left inset) SEM image of an as-grown 1.5 mm array of CNTs by sequentially feeding nickelocene and ferrocene each for 30 minutes. Five small TEM insets show catalyst particles following the length of a CNT (all in same magnification).

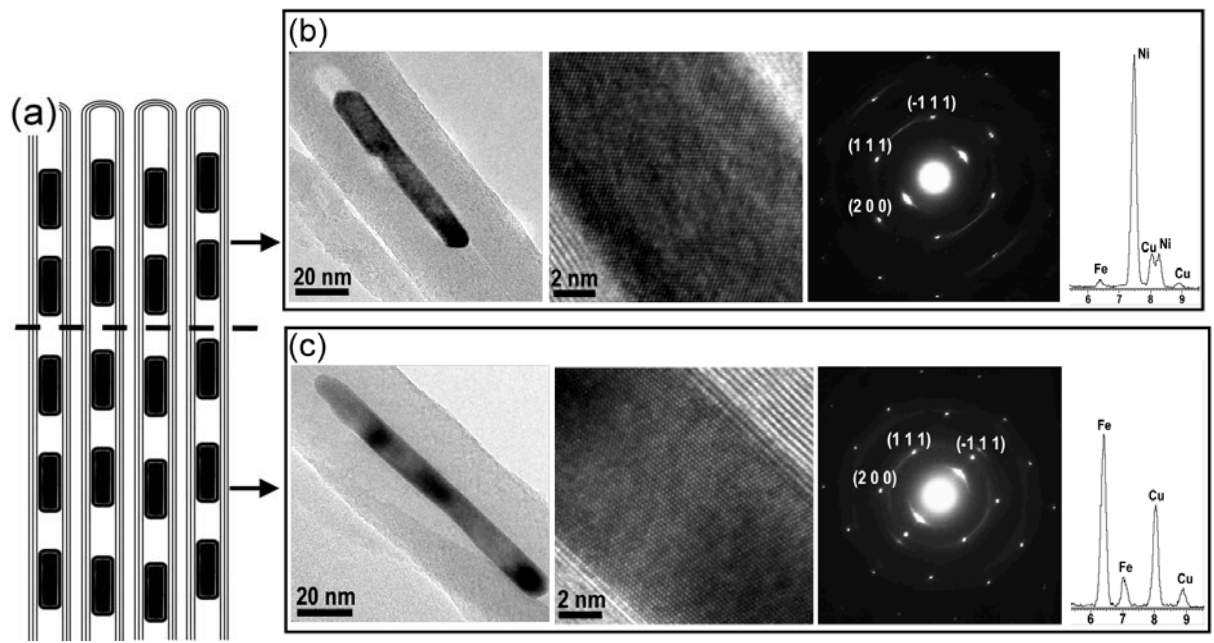

**Figure 2.** Tracking the distribution of catalyst materials. (a) Scanning was performed throughout the CNT array. (b) Ni particles distributed in the top portion of CNTs, from left to right showing low and high magnification TEM images, electron diffraction, and EDS on the particle (unit of horizontal coordinate in EDS is keV). The small Fe peak comes from the impurity of nickelocene, which was also detected in as-ordered nickelocene powders. Cu peaks are from the copper grid. (c) Fe particles found in the lower portion, with no trace of Ni detected by EDS. The dashed line in (a) represents the interface where the catalyst content changed from Ni to Fe.

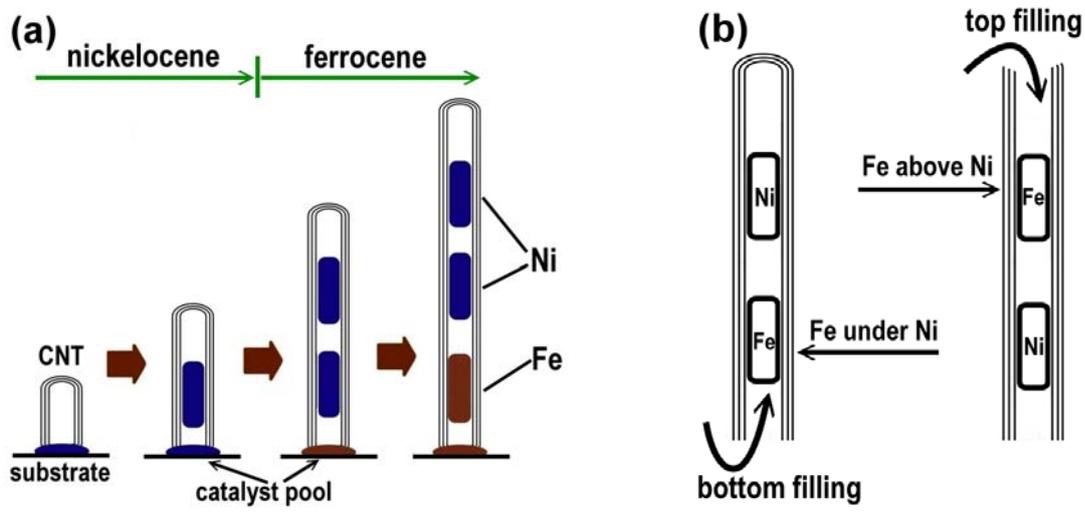

**Figure 3.** (a) Illustration of sequential feeding of nickelocene and ferrocene during CVD, and encapsulation of catalyst particles in CNTs. (b) Schematic of two configurations obtained from different filling mechanisms. The configuration is Fe under Ni if they enter the CNTs from the bottom, and Fe above Ni for top filling.

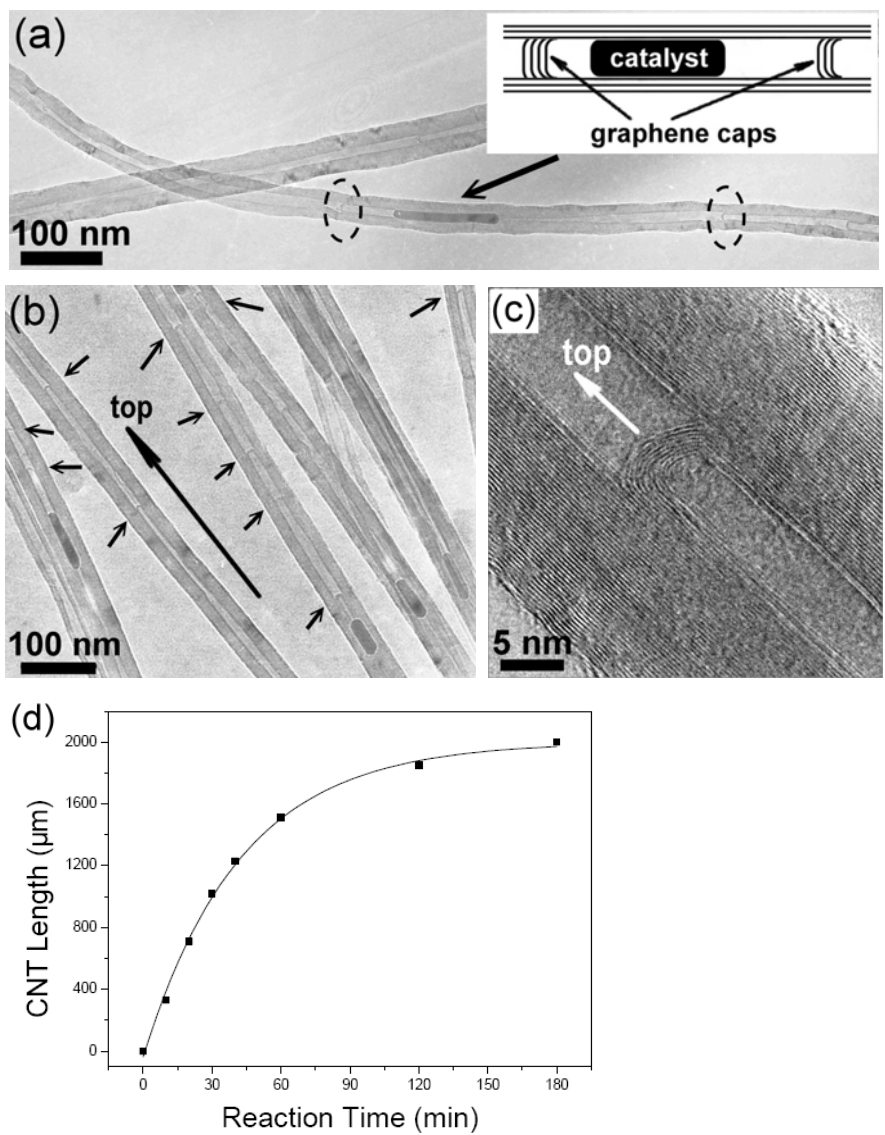

**Figure 4.** (a) A catalyst particle (see arrow) housed in a seal chamber between graphene caps (in dashed circles) at both sides. (b) (c) TEM images showing a large amount of graphene caps present in CNTs, oriented along the growth direction toward the top of film. (d) Plot of CNT length versus CVD time, showing reduced growth rate after 60 minutes.